\documentclass[conference]{IEEEtran}
\IEEEoverridecommandlockouts
\usepackage{cite}
\usepackage{algorithmic}
\usepackage{graphicx}
\usepackage{textcomp}
\usepackage{xcolor}

\usepackage[misc]{ifsym}
\usepackage{amsmath}
\usepackage{amssymb}
\usepackage{enumerate}
\usepackage{hyperref}
\usepackage{mathrsfs}
\usepackage{bm}
\usepackage{xspace}
\usepackage{graphicx}
\usepackage{float} 
\usepackage{url}
\usepackage{subcaption}
\usepackage{multirow}
\usepackage{tabularray}

\def\BibTeX{{\rm B\kern-.05em{\sc i\kern-.025em b}\kern-.08em
    T\kern-.1667em\lower.7ex\hbox{E}\kern-.125emX}}
\begin{document}

\title{KLDD: Kalman Filter based Linear Deformable Diffusion Model in Retinal Image Segmentation}


\author{\IEEEauthorblockN{Zhihao Zhao}
\IEEEauthorblockA{\textit{Technical University of Munich }\\
Munich, Germany \\
zhihao.zhao@tum.de}
\and
\IEEEauthorblockN{Yinzheng Zhao}
\IEEEauthorblockA{\textit{ Technical University of Munich }\\
Munich, Germany \\
yinzheng.zhao@tum.de}
\and
\IEEEauthorblockN{Junjie Yang}
\IEEEauthorblockA{\textit{ Technical University of Munich }\\
Munich, Germany \\
junjie.yang@tum.de}
\and
\IEEEauthorblockN{Kai Huang}
\IEEEauthorblockA{\textit{ Sun Yat-Sen University }\\
Guangzhou, China \\
huangk36@mail.sysu.edu.cn}
\and
\IEEEauthorblockN{Nassir Navab}
\IEEEauthorblockA{\textit {Technical University of Munich }\\
Munich, Germany \\
nassir.navab@tum.de}
\and
\IEEEauthorblockN{M.Ali Nasseri}
\IEEEauthorblockA{\textit{Technical University of Munich}\\
Munich, Germany \\
ali.nasseri@mri.tum.de}
}
\maketitle

\begin{abstract}
AI-based vascular segmentation is becoming increasingly common in enhancing the screening and treatment of ophthalmic diseases. Deep learning structures based on U-Net have achieved relatively good performance in vascular segmentation. However, small blood vessels and capillaries tend to be lost during segmentation when passed through the traditional U-Net downsampling module. To address this gap, this paper proposes a novel Kalman filter based Linear Deformable Diffusion (KLDD) model for retinal vessel segmentation. Our model employs a diffusion process that iteratively refines the segmentation, leveraging the flexible receptive fields of deformable convolutions in feature extraction modules to adapt to the detailed tubular vascular structures.
More specifically, we first employ a feature extractor with linear deformable convolution to capture vascular structure information form the input images. To better optimize the coordinate positions of deformable convolution, we employ the Kalman filter to enhance the perception of vascular structures in linear deformable convolution.
Subsequently, the features of the vascular structures extracted are utilized as a conditioning element within a diffusion model by the Cross-Attention Aggregation module (CAAM) and the Channel-wise Soft Attention module (CSAM). These aggregations are designed to enhance the diffusion model's capability to generate vascular structures. 
Experiments are evaluated on retinal fundus image datasets (DRIVE, CHASE\_DB1) as well as the 3mm and 6mm of the OCTA-500 dataset, and the results show that the diffusion model proposed in this paper outperforms other methods.
\end{abstract}


\section{Introduction}
\label{sec:introduction}
Retinal  vascular segmentation is a fundamental process in modern ophthalmology, playing a critical role in the diagnosis and monitoring of various ocular and systemic diseases. The complicated structure of blood vessels in the retina offers valuable insights, with abnormalities often serving as early indicators of diseases such as diabetic retinopathy, glaucoma, and age-related macular degeneration 
 \cite{bek2017diameter,zhao2023label,toulouie2022relationship}. If detected early, these conditions can be managed more effectively, highlighting the importance of accurate and detailed vascular segmentation. Moreover, the vascular structure in the retina is unique to each individual, making it a potential biomarker for biometric identification \cite{frost2013retinal,zhao2023unobtrusive}. The complexity of the retinal vascular system, characterized by varying vessel sizes, poses a significant challenge. Traditional methods of analysis, reliant on manual inspection or basic imaging techniques, are not only time-consuming but also prone to human error, emphasizing the need for advanced, automated segmentation techniques.

The evolution of retinal fundus vascular segmentation techniques has been marked by significant advancements, ranging from hand-crafted feature-based methods to  AI-driven approaches 
 \cite{jia2021learning}. Initially, hand-crafted methods such as matched filtering, vessel tracking, and morphological processing were employed \cite{odstrcilik2013retinal,yin2012retinal,hassan2015retinal}. While innovative at their inception, these techniques were limited in their ability to adapt to the variability in retinal images and were often ineffective in discerning finer vascular details. The advent of AI and machine learning introduced a new era, particularly with the emergence of deep learning models like convolutional neural networks (CNNs) \cite{chen2021retinal,oliveira2018retinal}. These AI-based methods have demonstrated superior performance in identifying and segmenting retinal vessels, owing to their ability to learn complex patterns and features from large datasets. However, challenges persist, especially in the segmentation of fragile vessels \cite{tang2020reswnet}. 
 Even with advanced techniques like U-net \cite{ronneberger2015u}, traditional CNN architectures often struggle to maintain the integrity of small vessels, resulting in incomplete or inaccurate segmentations.
 This limitation is predominantly due to the downsampling processes inherent in these networks, which can result in the loss of critical fine details essential for comprehensive vascular mapping.

Addressing the limitations of existing methods, our approach introduces a novel application of the diffusion model for retinal vascular segmentation. The primary advantage of our model is reflected in the diffusion model's ability to gradually reconstruct clear vascular structures from random noise. The powerful image representation capabilities of Diffusion models enable the network to better understand the correlations among pixels and to classify individual pixel points effectively.
Our model is structured into three primary components, with the diffusion model playing a fundamental role in generating the vascular structures. This process is essential for ensuring that the outcomes of vascular generation are controllable based on the input images.
To achieve this level of control, the network incorporates a retinal feature extractor that integrates linear deformable convolution, enhancing the model's ability to capture detailed vascular features from the input retinal images. In addition, Kalman filtering is employed to optimize the coordinate positioning of the deformable convolution's field of view, ensuring a more precise alignment with the actual vascular structures.
Lastly, to establish a stronger linkage between the input images and the generated vascular structures, we utilize the cross-attention aggregation module (CAAM) and the channel-wise soft attention module (CSAM). These modules are designed to enhance the relationship between the control images and the generated images, thereby ensuring that the generated vascular structures are not only accurate but also closely aligned with the specifics of the input retinal images. These approaches ensure a more detailed and accurate representation of the retinal vasculature, particularly by capturing the nuances of smaller vessels. 

\section{Related work}
\label{sec:related_work}


\subsection{Deformable Convolution in Segmentation}

Deformable convolution \cite{dai2017deformable}, designed to dynamically adjust to the geometric variations of input data, has emerged as a key research area in image segmentation. This interest is spurred by its proficiency in managing images with irregular shapes and structures. Gurita et al. integrated deformable convolution into segmentation frameworks, significantly enhancing the delineation accuracy of boundaries within medical images \cite{gurita2021image}. Furthermore, Yang et al. introduced a novel modulated deformable convolution. This approach incorporates a modulation mechanism, meticulously refining the sampling locations in a dynamic manner to bolster the model's capacity for capturing detailed structural details \cite{yang2022dcu}. In their paper \cite{qi2023dynamic}, Qi et al. proposed a dynamic snake convolution that accurately captures the features of tubular structures by adaptively focusing on slender and tortuous local structures.

\subsection{Diffusion Model in Segmentation}

Denoising Diffusion Probabilistic Models (DDPM) constitute an innovative category of generative models designed to acquire the capability of transforming a noise distribution into the distribution of data samples. \cite{ho2020denoising,huang2022draw,10.1145/3581783.3613857}.  In the field of image segmentation, diffusion models are uniquely employed to directly model or enhance the distribution of segmented images, leveraging their inherent generative power to refine segmentation outcomes derived through alternate methods \cite{wolleb2022diffusion,ottl2024analysing,jiang2023diffused}. 
Amit et al. \cite{amit2021segdiff} introduced a new approach to end-to-end segmentation using diffusion. They achieved this by integrating the information from the input image with the current segmentation map estimate through the aggregation of outputs from dual encoders. Subsequently, this integrated data is processed through diffusion models equipped with additional encoding layers and decoders, facilitating the iterative refinement of the segmentation maps.
Wu et al. have further advanced the field by proposing a unique application of diffusion models for segmenting medical images within the frequency domain \cite{wu2024medsegdiff1}. This approach underscores the profound advantages of diffusion models in medical imaging.

\begin{figure*}[ht] \centering
    \includegraphics[width=0.94\textwidth]{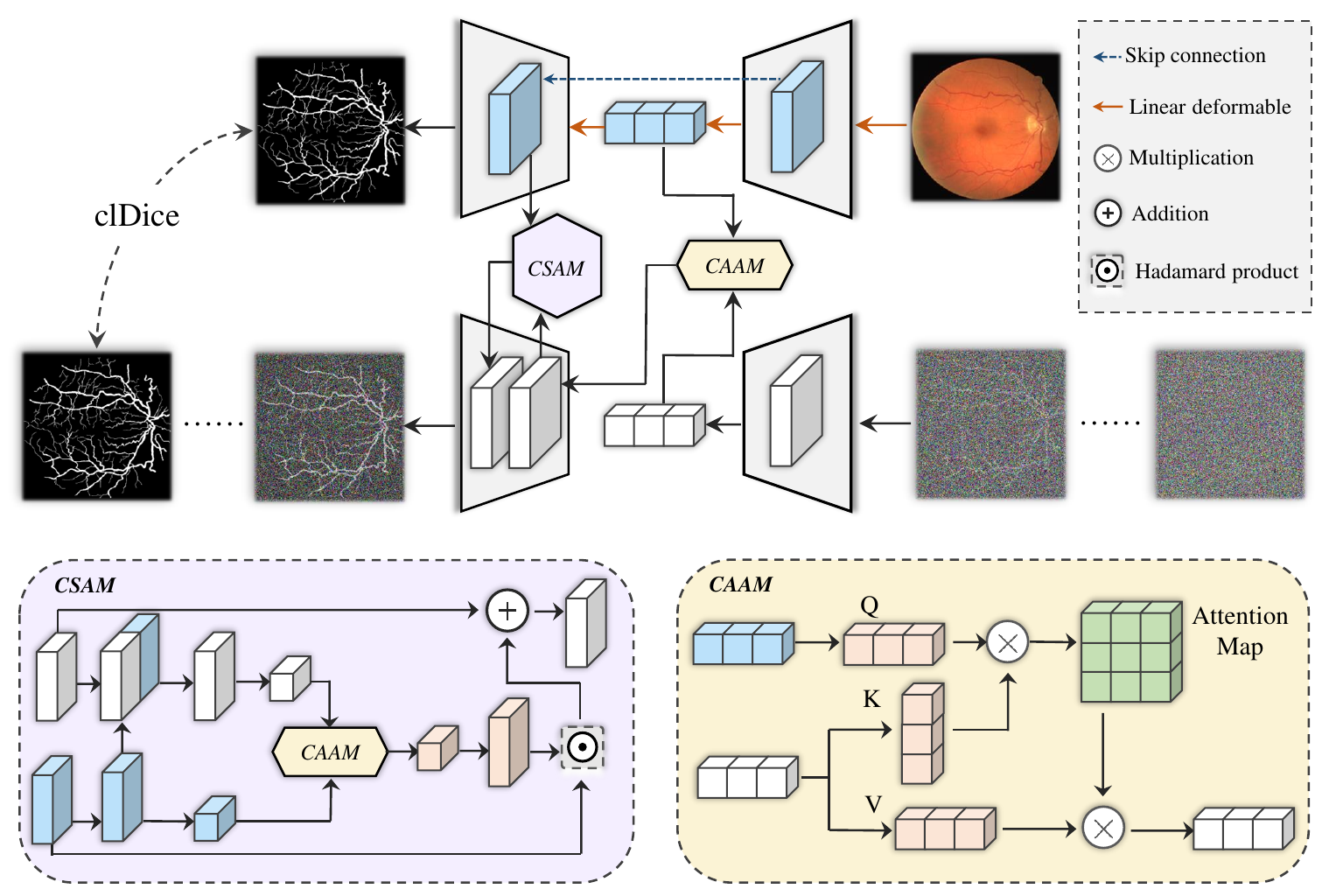}
    \caption{ Overview of our proposed KLDD model. The main structure of the model is based on diffusion processes, supplemented by an additional feature extractor designed to guide the generation of specific vascular structures. Within this extractor, a linear deformable convolution is applied to gather information specific to the vascular regions. The two feature aggregation modules, referred to as CAAM and CSAM, are mainly employed to aggregate the features extracted from the input image with those from the denoising module of the diffusion model.}
    \label{fig:framework}
\end{figure*}
\section{Methodology}
\label{sec:methodology}

\subsection{Overview of the Proposed Network}

Figure~\ref{fig:framework} illustrates the principal architecture of our model: a diffusion process that generates vascular structures by applying noise and subsequently denoising. To ensure that the synthesized vascular structures adhere to the constraints imposed by the input image, we integrated an additional feature extraction network. Within this network, we introduced an innovative approach to capture vascular structural information through the use of a novel linear deformable convolution. Furthermore, to minimize the accumulation of positional errors in the field of view for linear deformable convolution, we employed Kalman filtering for iterative optimization of coordinate positions in the deformable offsets. To guarantee that the vascular structures generated by the diffusion process are controlled by the input image, we employed the cross-attention aggregation module (CAAM) to merge the compressed vectors from both the feature extraction module and the diffusion model's noise prediction. Additionally, we utilized a channel-wise soft attention module (CSAM) in the decoder. This mechanism is designed to adjust the channel-specific weights during the fusion of features from both the linear deformable output and the diffusion denoising process, enhancing the perception of vascular morphology during denoising.

\subsection{Diffusion Model for Vessel Genetation}
Our model is  designed following the diffusion model framework as outlined in \cite{ho2020denoising}, which integrates both forward and reverse processes. The forward process, represented by $q(x_{1:T}|x_{0})$, either through a Markov or non-Markov chain, systematically transitions the initial data distribution $x_{0}\sim q(x_{0})$ to a state of pure noise $x_{T}$. In contrast, the reverse process, symbolized as $p_{\theta}(x_{0:T})$, employs a progressive denoising strategy to revert the noise sequence back to the original data distribution. This model employs a U-Net  to predict $x_{t-1}$ from $x_{t}$ for each step $t\in\{1,\ldots,T\}$. During the training phase, with a known ground truth for $x_{t-1}$, the model is optimized using the Mean Squared Error (MSE) loss. In the sampling phase, the process commences with noise $x_{T}\sim{\mathcal{N}}(0,\mathbf{I})$, iteratively sampling for T steps to synthesize the final image $x_{0}$.
The forward process q is described by the formulation:
\begin{equation}
q(x_{1:T}|x_0)=\prod_{t=1}^Tq(x_t|x_{t-1}),
\end{equation}
During each iteration of the forward process, Gaussian noise is incorporated in accordance with:
\begin{equation}
q(x_t|x_{t-1})=N(x_t;\sqrt{1-\beta_t}x_{t-1},\beta_tI_{n\times n}),    
\end{equation}
where $\beta_{t}$ is a constant that determines the schedule for introducing noise, and $I_{n\times n}$ is the n-sized identity matrix. 

Doing this for $t$ steps, we can write
\begin{equation}
q(x_{t}|x_{0}):=\mathcal{N}(x_{t};\sqrt{\overline{\alpha}_{t}}x_{0},(1-\overline{\alpha}_{t})\mathbf{I}),    
\end{equation}

with $\alpha_t: = 1- \beta_t$ and $\overline {\alpha}_t: = \prod _{s= 1}^t\alpha_s.$ With the reparametrization trick, we can directly write $x_t$ as a function of $x_0{:}$

The reverse process $p_\mathrm{\theta}$ is learned by the model parameters $\theta$ and is given by

\begin{equation}
p_{\theta}(x_{t-1}|x_{t}):=\mathcal{N}\big(x_{t-1};\mu_{\theta}(x_{t},t),\Sigma_{\theta}(x_{t},t)\big).    
\end{equation}

we can then predict $x_{t-1}$ from $x_t$ with
\begin{equation}
\begin{cases}
x_{t-1}=\dfrac{1}{\sqrt{\alpha_{t}}}\Bigg(x_{t}-\dfrac{1-\alpha_{t}}{\sqrt{1-\overline{\alpha}_{t}}}\epsilon_{\theta}(x_{t},t)\Bigg)+\sigma_{t}\mathbf{z},\\\\
z\sim\mathcal{N}(0,\mathrm{I})
\end{cases}
\label{equ:xt_1}
\end{equation}
Where $\sigma_t$ represents the variance scheme that the model can learn, as proposed in \cite{nichol2021improved}. Equation ~\ref{equ:xt_1} illustrates that sampling incorporates a random component $\mathbf{z}$, leading to a stochastic sampling process. It is noted that $\epsilon_{\theta}$ is the U-Net trained, with input $x_t=\sqrt{\overline{\alpha}_t}x_0+\sqrt{1-\overline{\alpha}_t}\epsilon$. The noise scheme $\epsilon_\theta(x_t,t)$, to be subtracted from $x_t$ during sampling as specified by Equation~\ref{equ:xt_1}, must be learned by the model.

\subsection{Vascular Features Extracted Module}

Our module for the extraction of vascular structure features utilizes a UNet architecture integrated with diffusion noise prediction. Within this module, we replace standard convolutions with our proprietary linear deformable convolution. This  convolution structure  inspired by Deformable Convolution Networks (DCNs) \cite{dai2017deformable} and Dynamic Snake Convolution Networks (DSCNet) \cite{qi2023dynamic}. DCNs dynamically adjust their field of view by predicting offsets, thereby extending the convolutional filed of view to better capture object-relevant regions.
The approach iteratively computes the complete receptive field by estimating the discrepancy between the current and previous locations. This technique effectively maintains the continuity of linear configurations, but it also facilitates the cumulative accumulation of predictive inaccuracies. In our research, we employ a strategy derived from Kalman filter theory to mitigate the error accumulation in computing offset positions within the linear deformable convolution. 

\begin{figure}[h]
    \centering \includegraphics[width=.46\textwidth]{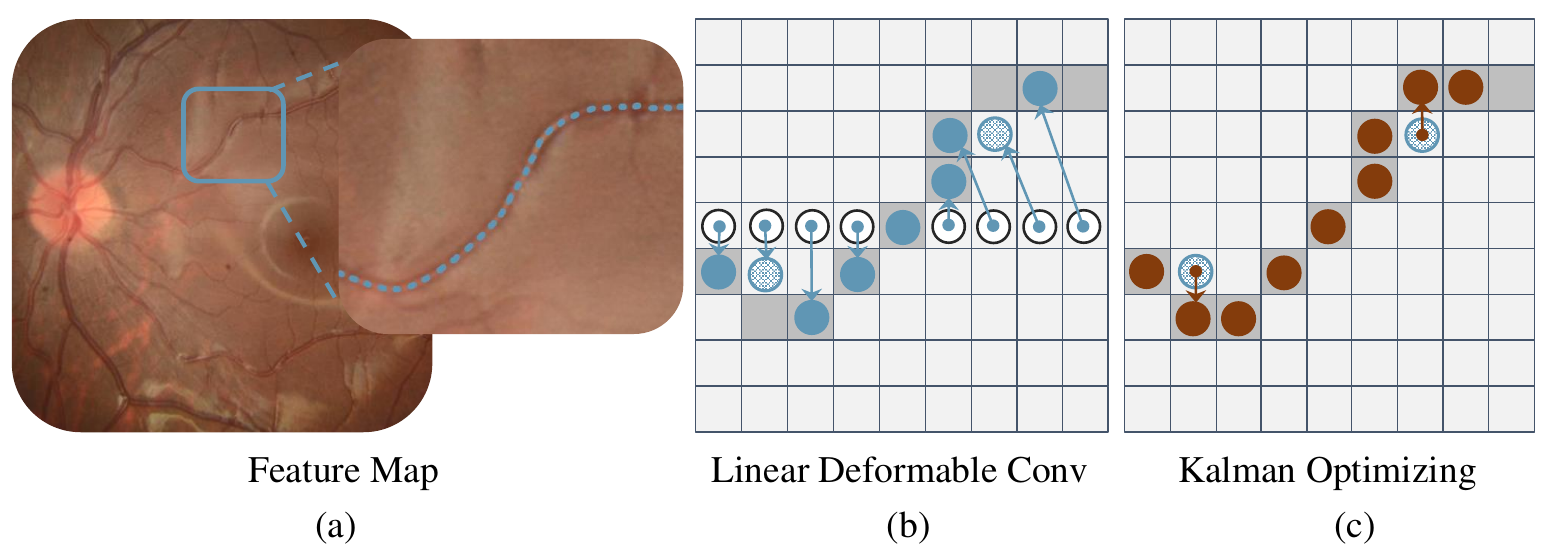}
    \caption{Figure (a) displays the feature map within the network. In Figure (b), the gray background illustrates the vascular area, while white circles symbolize standard one-dimensional convolution. Blue circles represent the positions of the field of view following linear deformable convolution, and blue grid circles indicate locations with aberrant offsets. In Figure (c), brown circles denote points that have been optimized using Kalman filter.}
    \label{fig:kalmanld}
\end{figure}

The linear deformation  module's key responsibility is learning the necessary offsets for adapting the convolution kernel's field of view. As depicted in Figure.~\ref{fig:kalmanld}(b), these learned offsets can shift the convolution kernel's initial field of view (white circles), to one that more closely aligns with the vascular structures (blue circles). 
As indicated by the blue grid circles in Figure.~\ref{fig:kalmanld}(b), some coordinates exhibit excessive offsets, necessitating correction for these excessive deviations. Therefore, Kalman filtering can be applied to optimize these deviations by weighting them with previous coordinate positions, leveraging historical information for optimization as shown in Figure.~\ref{fig:kalmanld}(c).
Kalman filter module iteratively reduces offset-induced errors, ensuring shifts more accurately reflect the true vascular structure. Following this adjustment, a linear convolution captures the adapted field of view, thereby allowing the entire module to adeptly feature-extract from vascular areas via deformable linear convolution.

In our work, one-dimensional linear convolution kernels sized $9\times1$ and $1\times9$ are employed. The discussion is focused solely on the horizontal coordinates, given that the vertical coordinates are identical. Each convolution kernel is denoted as $Ker=(x_{i\pm c})$, where $c=\{0,1,2,3,4\}$. DSCNet utilizes a learnable offset $\delta$ to predict the deviation in coordinates of the deformed convolution kernel $x_{i}=x_{i-1}+\delta_{i}$. Each new coordinate is the sum of the previous coordinate and the predicted deviation, resulting in cumulative errors.
To mitigate error accumulation in determining the positions within the offset of linear deformable convolution, our approach incorporates a Kalman filter based method. 
The Kalman filter offers an effective strategy for minimizing these errors by balancing current and past values \cite{welch1995introduction}. 
The optimization of linear deformable convolution by Kalman filtering is achieved by assigning a weight $K$ to the convolution kernel offset $\delta_{i}$. For a 1x9 convolution kernel, we start from the center of the kernel $x_0$, and then iteratively compute $x_i = (1-K_i)x_{i-1}+K_{i}(x_{i-1}+\delta_{i}) (i=1,2,3,4)$ using Kalman update Equation~\ref{equ:kal_update}.
Here, $K_{i}$ denotes the Kalman gain, computed iteratively based on Equation~\ref{equ:kal_update}, where $p_i$ signifies the estimate covariance and $r$ is a hyperparameter related to measurement errors from neural network outputs.

The hyperparameter $r$ is empirically set to $0.01$.
The initial values of $p_0$ and $x_0$ are set to $1$ and $0$, respectively. These parameters will be updated iteratively during the process. 
\begin{equation}
\begin{cases}
    K_{\mathrm{i}}=\frac{p_{i-1}}{p_{i-1}+r}\\
    x_i=(1-K_i)x_{i-1}+K_i(x_{i-1}+\delta_{i})=x_{i-1}+K_i\delta_{i}\\
    p_{i}=(1-K_i)p_{i-1}
\end{cases}   
\label{equ:kal_update}
\end{equation}
    
\subsection{Feature Aggregation Module by Attention}
The feature aggregation module is structured into two main components. The first component utilizes a cross-attention aggregation module (CAAM) in the encoder’s compressed feature space. This mechanism is designed to identify and emphasize the correlations between the features of the input image and the attributes obtained through the diffusion denoising process.
The second component entails the application of channel-wise soft attention module (CSAM) across decoder feature maps to learn the channel-specific weights between the feature maps of the input image and the diffusion denoising features, thereby guiding the focus on vascular regions throughout the diffusion denoising sequence. The inputs of the CAAM are two matrices of feature maps with same dimensions ($c\times h \times w$). These matrices are initially flattened and transformed into shape of $(h\times w)\times c$ . Subsequently, feature maps from both the input image and the diffusion denoising are converted into $QKV$ by a projection module, respectively. The product of matrices Q and K generates an attention map, which then multiplied by the V matrix, yields the aggregated outcome. Such mechanism enhances the diffusion denoising process's focus on the original image's vascular structures. Conversely, the CSAM initiates with the max pooling of the feature maps derived from the input image's linear deformable convolution,  encapsulating the vascular structure's control information into a single unit, thus reducing the feature maps from dimensions $h\times w \times c$ to $1\times 1\times c$. In parallel, the diffusion model's feature maps utilize average pooling, similarly downsample to $1\times 1\times c$, preserving only the fundamental information of each channel. Then, a cross-attention module evaluates the significance across various channels of the two vectors. Ultimately, the dimensions of the channel attention ($1\times 1\times c$), are expanded to match the dimensions of the original feature maps. Subsequently, weights derived from learned parameters are applied to both feature maps before amalgamating them into an updated feature maps.

\subsection{Loss Function}
In the segmentation of vascular structures, it is imperative not only to capture the details of vascular elements but also to ensure the overarching continuity within the vascular domain. To preserve topological continuity in the segmentation outcomes, a clDice loss is introduced, which leverages topological structure similarity as delineated in \cite{shit2021cldice}. We incorporate this topological similarity along with the standard noise prediction loss $\mathcal{L}_N$ to form our new loss. Topological similarity measurement is delineated by Equation~\ref{eq:dice}, where $V_L$ represents the ground truth of vascular segmentation, $V_P$ the predicted outcome, $S_L$ the skeleton derived from the ground truth, and $S_P$ the skeleton extracted from the predicted outcome. The topological accuracy $\mathrm{Tprec}(S_P,V_L)=\frac{|S_P\cap V_L|}{|S_P|}$ and topological sensitivity $\mathrm{Tsens}(S_L,V_P)=\frac{|S_L\cap V_P|}{|S_L|}$ are defined.
\begin{equation}
\begin{aligned}
   \mathcal{L}_{N} = & E_{x_0,\epsilon,t}[||\epsilon-\epsilon_\theta(\sqrt{\bar{\alpha}_t}x_0+\sqrt{1-\bar{\alpha}_t}\epsilon,t)||^2]\\
    \mathrm{clDice}(V_P,V_L)= & 2\times \frac{\mathrm{Tprec}(S_P,V_L)\times \mathrm{Tsens}(S_L,V_P)}{\mathrm{Tprec}(S_P,V_L)+\mathrm{Tsens}(S_L,V_P)}\\
\end{aligned}    
\label{eq:dice}
\end{equation}

Finally, we sum up the clDice and denoise loss functions together to form the loss function.
\begin{equation}
    \mathcal{L}_{oss}=\mathcal{L}_{N} + \mathcal{L}_{clDice}
    \label{eq:loss}
\end{equation}

\section{Experiment}
\label{sec:experiment}
\newsavebox\CBox
\def\textBF#1{\sbox\CBox{#1}\resizebox{\wd\CBox}{\ht\CBox}{\textbf{#1}}}

\subsection{Database}
All experiments are carried out on publicly available ophthalmic datasets. 
The DRIVE \cite{staal2004ridge} dataset contains 40 retinal images (584x565 pixels, 45° FOV), divided into training and test sets of 20 images each, with annotations from one or two experts. CHASE\_DB1 \cite{chasedb1} features 28 images (960x999 pixels, 30° FOV) from 14 children (one image per eye), each annotated by two experts. OCTA-500 \cite{li2024octa} comprises 500 OCTA projection images: 300 images with a 6mm×6mm field of view and 200 images with a 3mm×3mm field of view, all provided with corresponding annotations.

\subsection{Evaluation Metrics}
We calculate the area under the ROC curve (AUC) between the segmentation results and the ground truth. Additionally, we systematically evaluate other segmentation metrics, including accuracy: $Acc=(TP+TN)/(TP+TN+FP+FN)$, sensitivity: $Sen=TP/(TP+FN)$, specificity: $Spe=TN/(TN+FP)$, F1 score or DICE score: $F1=DICE=2TP/(2TP+FP+FN)$, and Intersection over Union: $IOU=(TP)/(TP+FP+FN)$).
\subsection{Implementation Details}
Our experimental setup utilizes a single NVIDIA RTX A5000 GPU with 24GB memory and is implemented using PyTorch. The learning rate is initially set to 1e-4, and for optimization, the Adam optimizer is employed with a weight decay of 1e-5. 
During the training phase, preprocessing of images is conducted as follows: color images are converted to grayscale and uniformly normalized. Subsequently, all images are first divided into patches of $256\times256$, and then during the inference stage, the segmentation results for each patch are reassembled.
The dataset is trained over 100 epochs. To mitigate overfitting, online data augmentation techniques such as horizontal and vertical flipping are employed, along with the addition of Gaussian noise using a 5x5 kernel.  

\begin{figure*}[ht] \centering
    \includegraphics[width=0.92\textwidth]{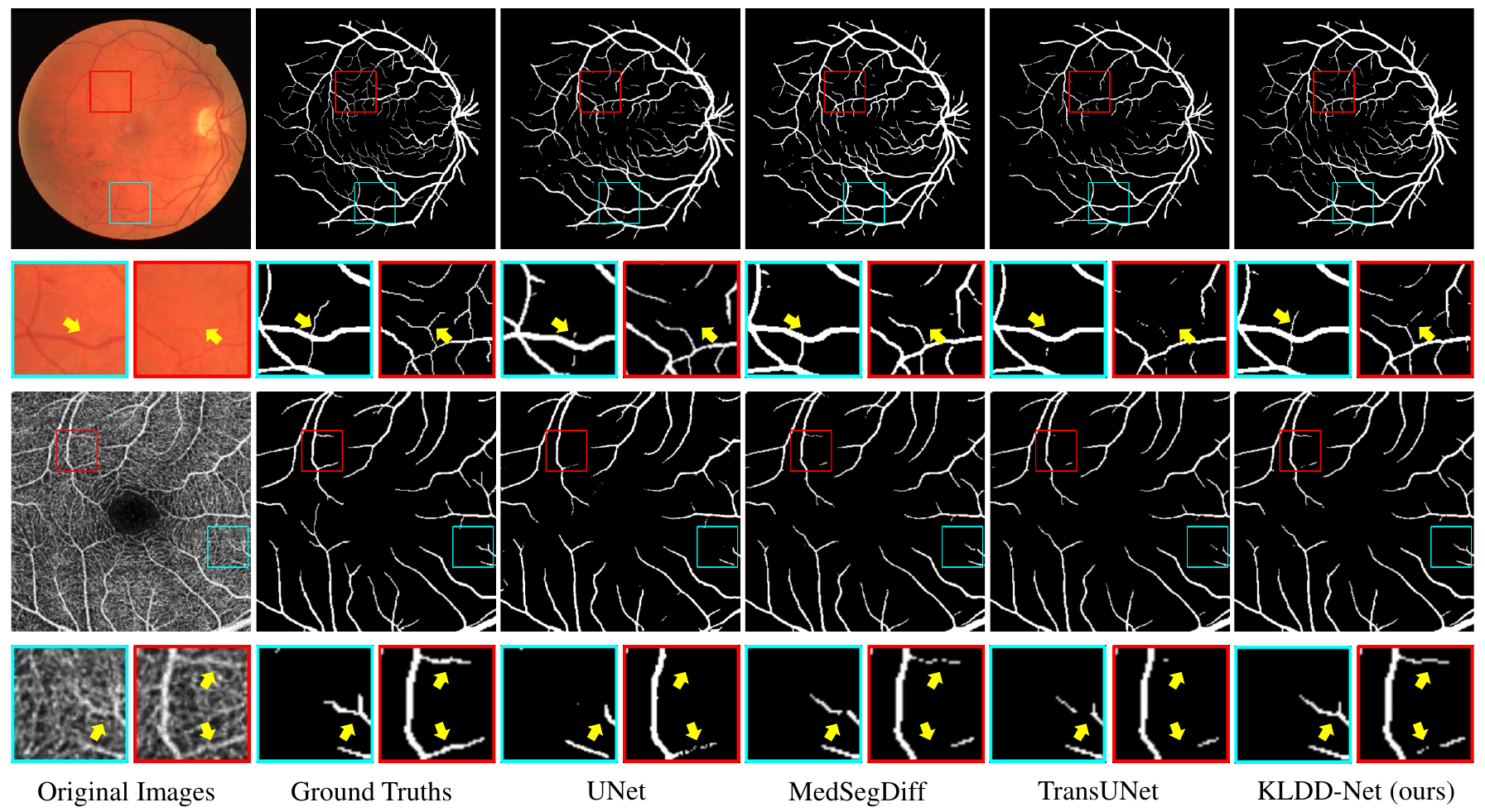}
    \caption{The visualization of segmentation results is organized as follows: the first and third rows show the segmentation outcomes of fundus and OCTA, respectively, while the second and fourth rows illustrate enlargements of the  cropped regions.}
    \label{fig:segmentresult}
\end{figure*}
\subsection{Results of Segmentation}
In the experiment section, we presented a comprehensive evaluation of our proposed model's performance in segmenting vascular structures from medical imaging datasets, specifically focusing on fundus images and Optical Coherence Tomography Angiography (OCTA) datasets. 
We compared with models from the UNet family, including UNet \cite{ronneberger2015u} and FR-UNet \cite{liu2022full}, along with PVT-GCASCADE \cite{rahman2024g} which employs attention mechanisms, and SwinUNETRV2 \cite{he2023swinunetr} and  TransUNet \cite{wang2020rvseg}, which incorporate transformer architectures. Additionally, DUnet \cite{jin2019dunet}, which utilizes deformable convolutions, and MedSegDiff \cite{wu2024medsegdiff1}, which integrates diffusion techniques, were also employed for benchmark testing.
Figure~\ref{fig:segmentresult} displays the segmentation results derived from both our model and conventional network architectures.  From the cropped sections in the illustration, we can observe that numerous delicate and barely visible vascular formations, crucial for precise diagnosis and commonly present in retinal images, are consistently overlooked or inadequately captured by traditional networks. Despite their effectiveness across various applications, compared networks lack the necessary sensitivity for identifying the smallest vessels, which are crucial indicators for the early detection of a range of ocular diseases. Conversely, our network exhibits remarkable capability in the identification and segmentation of small vessels, adeptly capturing complex vascular nuances frequently missed by conventional methods. This success is attributed to the intrinsic characteristics of our approach, which fundamentally utilizes a diffusion model for image generation guided by the features of the original images, thereby generating images with enhanced continuity. Furthermore, our adaptive linear deformable structure effectively seizes the features of vascular structures. The inclusion of CSAM and CAAM aids in enhancing feature aggregation, which in turn more effectively captures the relationship between the vascular structures in the input images and the noise-reduced feature maps generated through diffusion. 
\begin{table*}[th]
  \centering
  \caption{Comparison with the State-of-The-Art Methods on DRIVE and CHASE\_DB1 }
  \label{tab:drivechase}
  \resizebox{\textwidth}{!}{%
  \begin{tblr}{
    cell{1}{1} = {r=2}{},
    cell{1}{2} = {r=2}{},
    cell{1}{3} = {c=6}{c}, %
    cell{1}{9} = {c=6}{c},
    vline{2,3,9} = {1-14}{},
    hline{1,3,10,11} = {-}{},
    hline{2} = {3-15}{},
  }
  Methods         & Year  & DRIVE (\%) &  &       &       &       &       & CHASE\_DB1 (\%) &  &  &       &       &        \\
                  &       & Acc   & Sen   & Spe   & AUC   & Dice  & IOU   & Acc   & Sen   & Spe   & AUC   & Dice  & IOU    \\
  U-Net  \cite{ronneberger2015u}         & 2015  & 96.78 & 80.57 & 98.33 & 98.25 & 81.41 & 68.64 & 97.43 & 76.50 & \textBF{98.84} & 98.36 & 78.98 & 65.26 \\
  DUNet   \cite{jin2019dunet}       & 2019  & 96.81 & 78.31 & 98.50 & 98.26 & 81.14 & 68.26 & 97.38 & 83.52 & 98.37 & 98.74 & 80.16 & 66.84 \\
  TransUNet  \cite{chen2021transunet}     & 2021  & 96.62 & 79.06 & 98.31 & 97.74 & 80.39 & 67.21 & 97.30 & 83.84 & 98.20 & 98.48 & 79.64 & 66.17 \\
  SwinUNETRV2   \cite{he2023swinunetr}      & 2023  & 96.98 & 79.83 & 98.38 & 97.86 & 79.96 & 66.61 & 97.46 & 82.83 & 98.38 & 98.76    & 79.37 & 65.80 \\
  FR-UNet   \cite{liu2022full}      & 2022  & 97.05 & \textBF{83.56} & 98.37 & \textBF{98.89} & 83.16 & 71.20 & 97.48 & \textBF{87.98} & 98.14 & \textBF{99.13} & 81.51 & 68.82 \\
  PVT-GCASCADE  \cite{rahman2024g}  & 2023  & 96.89 & 83.00 & 98.22 & --    & 82.10 & 69.70 & 97.71 & 85.84 & 98.51 & --    & 82.51 & 70.24 \\
  MedSegDiff   \cite{wu2024medsegdiff1}   & 2024  & 97.37 & 80.12 & 98.78 & 98.17 & 82.14 & 69.69 & 97.86 & 84.37 & 98.72 & 98.87 & 82.56 & 70.30  \\
  KLDD(ours)      & 2024  & \textBF{97.55} & 80.58 & \textBF{98.98} & 98.29 & \textBF{83.65} & \textBF{71.90} & \textBF{98.04} & 86.28 & 98.78 & 99.01 & \textBF{83.85} & \textBF{72.19}           
  \end{tblr}
  }
  \end{table*}


Moreover, our  evaluation encompasses a range of metrics, including Area Under the Curve (AUC), Accuracy (ACC), and DICE coefficients, to conduct quantitative analysis. This analysis serves to underscore the enhanced performance of our proposed model relative to alternative approaches. Detailed quantitative comparisons among various models on the DRIVE and CHASE\_DB1 datasets are provided in Table~\ref{tab:drivechase}, illustrating the effectiveness of our method in segmenting vascular structures within retinal images captured through fundus photography.  The crucial role of Table~\ref{tab:drivechase} in underscoring our model's precision in identifying vascular structures, a critical aspect of retinal image analysis, is emphasized. Additionally, Table~\ref{tab:octa} outlines our model's segmentation capabilities on the OCTA dataset, further validating the proposed structure's versatility and robustness across different imaging modalities.

The empirical data in Tables~\ref{tab:drivechase}compellingly establish new  standards for accuracy (ACC) on the DRIVE, and CHASEDB1 datasets, achieving remarkable scores of 97.55\% and 98.04\%. The performance is further solidified by our method's superior results in DICE scores when compared against alternative methods across the DRIVE and OCTA datasets. However, the scope of our method's superiority is not limited to just ACC and DICE scores. A broader examination across additional evaluative metrics reveals our method's strength, showcasing a consistent and robust performance that transcends traditional evaluation paradigms.
Furthermore, the thorough analysis of our results underscores the methodological rigor and scientific inquiry underpinning our approach. By setting new benchmarks in ACC and DICE scores and extending the evaluation to encompass a broader range of metrics, our method demonstrates a holistic and nuanced understanding of the challenges inherent in retinal image analysis.

Within the analytical scope of the OCTA dataset, our investigation primarily focused on two key metrics: Accuracy (ACC) and the DICE coefficient. Remarkably, the accuracy achieved on the OCTA\_3mm and OCTA\_6mm datasets stood at  98.92\% and 98.23\%, respectively. Furthermore, the DICE scores recorded for the OCTA\_3mm and OCTA\_6mm datasets were 91.28\% and 89.08\%, respectively, demonstrating our method's superior capability in precise segmentation.
This exploration into the OCTA dataset's complexities not only affirms our model's superior performance but also enriches our understanding of its broad applicability and the technological strides we have achieved.


\begin{table}[h]
    \centering
    \caption{ Quantification results on OCTA}
    \label{tab:octa}
    \resizebox{0.46\textwidth}{!}{%
    \begin{tblr}{
      cell{1}{1} = {r=2}{},
      cell{1}{2} = {c=2}{c}, %
      cell{1}{4} = {c=2}{c},
      vline{2,4} = {-}{},
      hline{1,3,8,9} = {-}{},
      hline{2} = {2-6}{},
    }
    Methods        & OCTA\_3mm (\%)  &     & OCTA\_6mm (\%) &      \\
                   & Acc    & Dice   & Acc   & Dice    \\
    U-Net \cite{ronneberger2015u}      & 95.45 & 88.35 & 95.21 & 85.03 \\
    DUNet \cite{jin2019dunet}    & 97.52 & 88.22 & 96.73 & 87.70 \\
    TransUNet \cite{chen2021transunet}    & 96.32 & 90.89 & 97.42 & 88.67 \\
    FR-UNet \cite{liu2022full}     & 98.84 & 91.15 & 98.02 & 88.85 \\
    MedSegDiff \cite{wu2024medsegdiff1}      & 98.87 & 90.87 & 98.18      & 88.68      \\
    KLDD(ours) & \textBF{98.92} & \textBF{91.28} & \textBF{98.23} & \textBF{89.08}           
    \end{tblr}
    }
\end{table}

\begin{figure}[ht]
    \centering
    \includegraphics[width=0.46\textwidth]{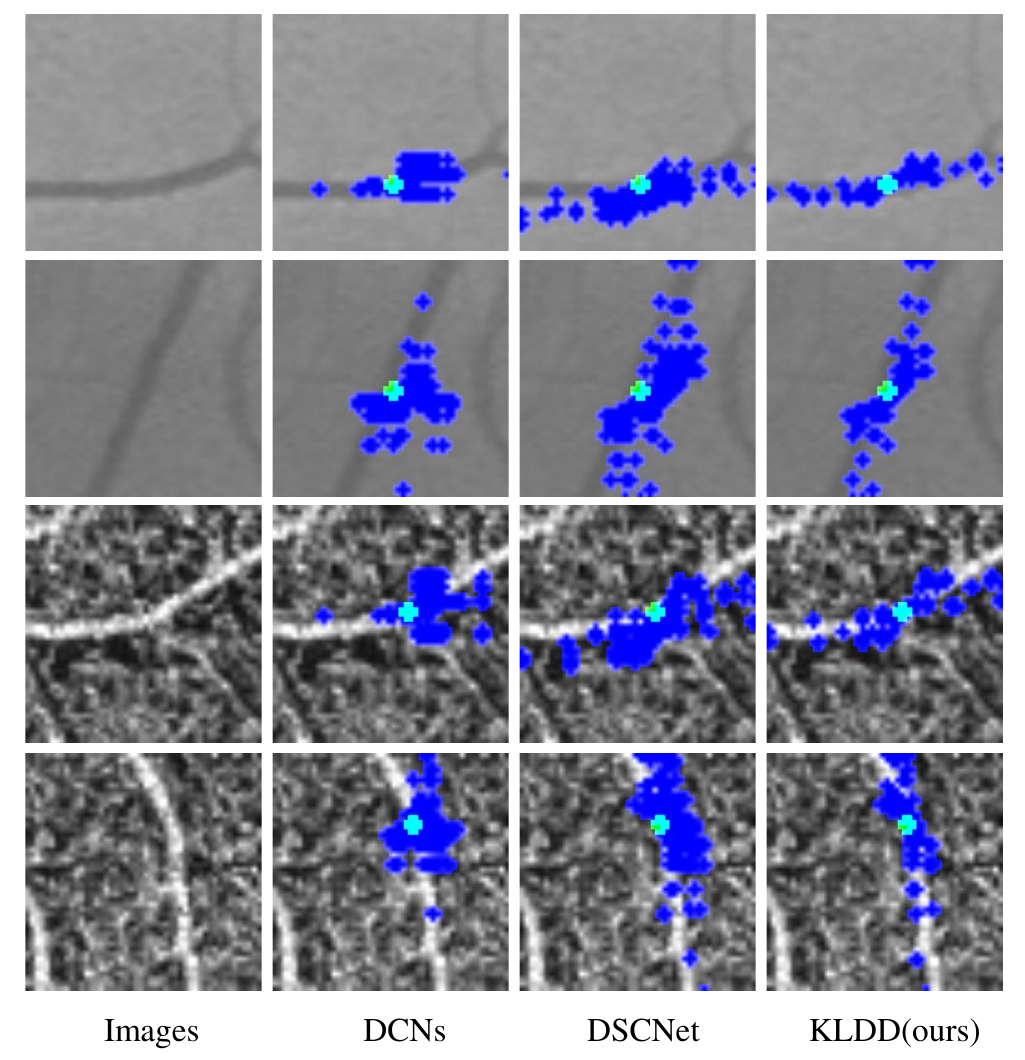}
    \caption{The first and second rows illustrate fundus images, specifically showcasing the convolutional fields of view for horizontal and vertical vascular structures under various deformable convolution settings. Similarly, the third and fourth rows display the visualized fields of view for deformable convolution applied to OCTA images.}
    \label{fig:visdcn}
\end{figure}

\subsection{Structural Continuity Preservation}
To assess our method's effectiveness in preserving the overall continuity of vascular structures, we employed the Centerline Dice (clDICE) metric to evaluate the topological continuity of tubular structures. As illustrated in Table~\ref{tab:exp:continuous}, our method consistently surpasses others in sustaining vascular continuity across all tested datasets. This superior performance suggests that KLDD is particularly adept at maintaining the overall continuity of vascular structures, a crucial aspect for dependable medical image analysis. In the ablation study section, we will further explore the influence of our model on vascular continuity.

    \begin{table}[th]
            \centering
            \caption{Structural Continuity Preservation evaluation using clDICE metric }
            \label{tab:exp:continuous}
            \resizebox{.48\textwidth}{!}{%
            \begin{tblr}{
              vline{2,3,4,5} = {-}{},
              hline{1,2,7,8} = {-}{},
            }
            Methods         & DRIVE(\%)  & CHASE\_DB1(\%)    & OCTA3mm(\%) &OCTA6mm(\%) \\
            U-Net \cite{ronneberger2015u}           & 75.71      & 78.92    & 86.98        & 86.74   \\
            DUNet \cite{jin2019dunet}          & 76.33      & 81.62    & 89.30        & 87.14   \\
            TransUNet \cite{chen2021transunet}      & 77.67      & 81.31    & 89.56        & 87.98   \\
            FR-UNet \cite{liu2022full}        & 82.39      & 84.20    & 92.98        & 90.57   \\
            MedSegDiff \cite{wu2024medsegdiff1}         & 80.57      & 83.53    & 90.49        & 87.91   \\
            KLDD(ours)    & \textBF{83.07} & \textBF{85.06} & \textBF{94.13} & \textBF{92.12}
            \end{tblr}
            }
    \end{table}

\subsection{Ablation Studies}

\begin{table*}[h]
  \centering
  \caption{Ablation study of the proposed modules on different datasets}
  \label{tab:ablation}
  \resizebox{\textwidth}{!}{%
  \begin{tblr}{
    cell{1}{1} = {r=2}{},
    cell{1}{2} = {c=3}{c}, %
    cell{1}{5} = {c=3}{c}, %
    cell{1}{8} = {c=3}{c}, %
    cell{1}{11} = {c=3}{c}, %
    vline{2,5,8,11} = {1-6}{},
    hline{1,3,7} = {-}{},
    hline{2} = {2-17}{},
  }
  Methods                       & DRIVE (\%)  & &       & CHASE\_DB1 (\%) & &   &  OCTA\_3mm (\%) &    && OCTA\_6mm (\%) && \\
                                & Acc   & Dice  &clDice & Acc   & Dice  &clDice & Acc    & Dice &clDice & Acc   & Dice   & clDice \\
  
  Baseline                      & 96.98 & 79.96 & 78.72 & 97.46 & 79.37 & 81.54 & 97.52 & 88.22 & 89.96 & 96.73 & 87.70 & 88.73 \\
  Baseline+LD                   & 97.20 & 81.94 & 78.80 & 97.77 & 82.03 & 81.67 & 98.76 & 90.57 & 89.94 & 97.79 & 88.05 & 89.02 \\
  Baseline+LD+Kalman            & 97.25 & 82.12 & 80.32 & 97.83 & 82.51 & 82.84 & 98.84 & 90.89 & 91.26 & 98.09 & 88.64 & 89.88 \\
  Baseline+CSAM+CAAM+LD+Kalman  & 97.55 & 83.65 & 83.07 & 98.04 & 83.85 & 85.06 & 98.92 & 91.28 & 94.13 & 98.23 & 89.08 & 92.12 \\
  \end{tblr}
  }
  \end{table*}

To evaluate the effectiveness of each module proposed in our study, we conducted ablation experiments on individual submodules, the results of which are displayed in Table~\ref{tab:ablation}. This ablation study assessed the contributions of the proposed modules integrated into a DDPM baseline. Integrating the linear deformable module alone resulted in a significant improvement in segmentation performance, increasing the Accuracy (ACC) by nearly 1\% and the DICE score by an average of 2\%. 
The Kalman filtering module has also contributed to an improvement in segmentation accuracy, particularly reflecting an average increase of nearly 1\% in the vascular continuity metric (clDICE). We also visualized the receptive field of our linear deformable convolution in Figure~\ref{fig:visdcn}, which shows our LD module provides a tighter fit to the vascular structure.

In addition, the inclusion of CSAM and CAAM for feature aggregation has increased accuracy by 0.5\%. More importantly, the integration of the feature aggregation modules has resulted in an enhancement of more than 2.5\% in the vascular continuity metric for segmentation outcomes. This indicates that the feature aggregation modules CSAM and CAAM are more effective globally in capturing the overall structural information of the vessels.

\section{Conclusions}
In this paper, we introduce a network called KLDD-Net, which is structured around a diffusion model as its primary backbone. This network integrates a specialized feature extraction module that incorporates linear deformable convolution, with Kalman filtering employed to optimize the convolution's field of view. This module plays a crucial role in extracting salient features from the input images. These features are subsequently utilized as control information, integrated into the diffusion denoising module.
The embedding vectors generated by the encoder within the feature extraction module and those from the diffusion denoising module are aggregated through a cross-attention mechanism. This integration allows the diffusion denoising module to effectively incorporate and process information from the input images within the space of embedding vectors.



\bibliographystyle{IEEEtran}
\bibliography{main}

\end{document}